\documentstyle[preprint,floats,pra,aps]{revtex}

\tightenlines 

\begin{document}
\draft

IPT: October, 1999/ Abstract for APS Meeting/ \\

\bigskip

\bigskip

\begin{center}
{\large \bf A biological junction with quantum-like characteristics }\\

\bigskip

{\large Alex A. Samoletov} \\

{\sl NASU - Institute for Physics and Technology\\
72 Luxembourg Str., 340114 Donetsk, Ukraine\\
samolet@kinetic.ac.donetsk.ua}\\
\end{center}
\bigskip

A model of chemical synapse as an electric junction is proposed. Estimations
and  analysis of the model show that the junction has unique physical
characteristics reminding the Josephson junction. The basic assumption is
made that the electric coupling across the synaptic gap is indirectly
provided by means of approximately quantized portions of a chemical
mediator, each the portion is content of a synaptic bubble. We suppose that
effective quantum of charge is $q$, $|q|\gg |e|$. The synapse
characteristics are dominated by electrostatic energy, $Q^2/2C$, $Q=qN$, $%
N=0,1,2...$; where $C$ is electric capacity of membrane. Estimations show
that the integer-valued character of $N$ must be explicitly taken into
account. The consistent theory of the junction is constructed on the basis
of operator realization of number-phase canonical pair in the Hardy space.
The charge passing from one side of the junction to other is described by
the Toeplitz operators. The synapse state space is constructed explicitly.
The unique physics of the model is investigated in detail. We do not exclude
the possibility that the model is prototype of a molecular electronics
device.

\newpage
IPT: March, 1999/ Conference Paper/ ICSSUR'99/ 
\bigskip
\bigskip
\begin{center}
{\large \bf A model of mesoscopic junction:\\
The benefits of number-phase operators\\
(application in biophysics)}\\
\bigskip
{Alex A.~Samoletov}\\
{\small \sl NASU -- Institute for Physics and Technology\\
72 Luxembourg Str., 340114 Donetsk, Ukraine\\
samolet@kinetic.ac.donetsk.ua}
\end{center}

\begin{quote}
{\small In this talk we present a primary approach to the physical modeling
of a chemical synapse as the quantum-like junction in a nearly closed loop
of a self-synapsing neuron. We consider such kind of biological systems as a
candidate for detecting and processsing of microwave radiation. Explicit
realization of the state space and the junction Hamiltonian operator are
constructed consistently.} 
\end{quote}

\bigskip 

The influence of electromagnetic waves, especially of microwave radiation,
on biological systems has attracted much interest for a long time. In
particular, active areas of research have been the study of the results of
radio- and microwave radiation effect in biological populations, or the
possible influence of usual household electrical appliances of the new
generation on humans and others living systems, or an accident prevention in
industry. These investigations are of interest mainly from a
phenomenological viewpoint since the works on cellular or molecular level
are rare up to now. They are also of interest from the point of view of
understanding how the electromagnetic waves of different frequency and power
affect biological systems. By the way, there were the rumors that in the
former USSR the building of the USA Embassy in Moscow has been exposed to
low-level microwave radiation during of many years. On the other hand, it
seems that the interest of physicists in this subject is concentrated also
in the field of macroscopic electrodynamics phenomenology, being quite less
rich that electrodynamics of complex nonequilibrium cellular- and
molecular-level biological systems.

At the present time, at least in Donetsk, Ukraine, there is the official
medical institution for therapy and research, Hospital ''Sitko''-- MRT
(microwave therapy), where treat patients with extremely low-level microwave
radiation by the special procedures. Thus, some aspects of the microwave
therapy are clinically tested.

The present work was originally motivated by the desire to understand a
possible mechanism of the effect of a low-level microwave radiation on
biological objects such as humans (or rats, for example),-- and on a
cellular level, beyond the phenomenology. It is in connection with the
microwave therapy. On the other hand, the desire was inspired by the recent
works in the field of neurophysiology. Namely, our model accepts a
hypothesis which is based on the existence of in a sense mesoscopic
self-synapsing neurons, neurological loops with chemical synapses. The
experimental evidence that such the loops are really existing has been
recently reported $\left[ 1\right] $ (concerning young rats).

In the present talk, we set up the physical model and the basic theoretical
framework we shall need for the study of detection and processing of
extremely low-level microwave radiation by the biological systems on the
cellular level. The results we present here will be used in the following
works.

\medskip

Before starting to work out the physical model and its relations with the
biological system, it may be worth getting some idea of what it is about, in
particular what is a synapse and what is a junction. Of course, in a
schematic, without any details, way.

Above all, let us to give some preliminary idea about the neurological
system under consideration. A typical neuron has about $10^3\div 10^4$
synapses. Self-synapsing neuron is a nearly closed, circular loop of
electrically excitable (neuron membrane) conducting material, the biological
loop of axon and its dendrite, of in a sense mesoscopic size (to be defined
below). The thickness of a neuron membrane is about $50$\AA\ and the
electric capacity of membrane is about $1\mu F/cm^2$. The loop contains a
gap. It is the gap of a synapse (usually $\sim 10^2$\AA ). We suppose that
it is a chemical synapse. It means that the electrical coupling across the
synaptic gap is indirectly provided by means of quantized portions of
chemical mediator. Each the portion, the content of a synaptic bubble,
contains about $10^3$ molecules of a chemical mediator. We suppose that an
effective ''quantum'' of charge in synapse is about $10^3e$. For further
relevant and more detailed information concerning neurons, synapses and all
that we refer the reader to $\left[ 2\right] $.

\medskip

Then, the notion 'mesoscopic junction' is generic for a wide class of the
physical systems. The tunnel junction is a prototype the junction class, of
course the last has the matter far beyond this, that illustrates the
relevant physical phenomena and the corresponding theoretical prolegomena.
Physical systems like mesoscopic junctions are widely established in current
physical literature.

Here, beginning with the simple physical model of a complex biological
junction, we construct the consistent quantum-like theory of mesoscopic
junctions including the explicit realizations of the model state space and
the Hamiltonian operator as an operator in this space, and in what follows
we describe in this frame the corresponding dominant physical effects
together with application to the biological system, such as a self-synapsing
neuron, which is treated as a candidate on the role of detector and
processor of microwave radiation in biological systems.

\medskip \medskip

The physical model consists in the following. A self-synapsing neuron system
is idealized as a circular loop of electrically conducting material. The
synaptic gap in this loop is modeled as a junction of relatively small
capacity $C$ . Really it is a system of two membrane capacities connected by
physiological solvent; however, it is easy to find the arguments that we can
replace this system by single effective capacity (e.g., the resistance
across a synapse is dominated by the membranes). But, on the other hand, the
charge carriers in this junction are ''quantized'' due to chemical nature of
the synapse. It means that an effective elementary charge $Q$ may be
considerably greater then the charge of electron and every additional charge 
$Q$ will change electrostatic energy on the junction substantially. Under
the such conditions the role of charge energy on the junction increase and
we must to take the quantum-like nature of the effective charge into
account. All that is the first part of the system mesoscopicity condition
mentioned above. The second one is the geometric size (radius) of the loop.
This second aspect is connected with the fact that magnetic fields penetrate
biological tissue much more effectively then electric fields and thus the
geometric size of the loop is directly connected with magnetic flux through
the loop, and will dominate in detection and processing of microwave
radiation, for example, by means of a depolarization of the membrane and an
induced exit of a mediator into the synaptic gap. But it is the topic of
another paper.

\medskip \medskip

In order that the model be more mathematically formulated, it is
sufficiently to define the character of relevant macrovariables. We set the
number $N$ of $Q$ carriers as the characteristic macrovariable. Further, we
assume that in respect of the characteristic macrovariable an homogeneous
state on the junction is realized. And also we take into account the
discreteness (''quantization'' by $Q$) of a charge magnitude on the junction
explicitly. This implies that the relevant is setting as fundamental the
canonical pair of the action-angle (number-phase) operators realized on a
proper state space; we realize the state space as the Hardy space ${\bf H}^2$
(e.g., $\left[ 3\right] $).

In this point the principal from the theoretical point of view and crucial
for the theory question is arising: How much a wealth of material can be
extracted from the model to be restricted to the fact of discreteness of a
charge carriers and under conditions of a system mesoscopicity (e.g.,
concerning electric capacity, inductance or geometric size)? The answer to
this question give us the key to a lot of the problems.

Then, there are usual arguments that after a coarse-graining procedure a
quantum-like energy operator, the Hamiltonian operator $H,$ is a function of
the variable $N$ only. By the way, it is in perfect harmony with
Ginzbirg--Landau phenomenology. Indeed, if the canonical pair of operators $%
\left( N,\Phi \right) ,\ \left[ \Phi ,N\right] =i,$ is defined in the Hardy
space ${\bf H}^2$ (it is reasonable way) then using isometry ${\bf H}%
^2\rightarrow {\cal L}^2$ and the Wigner phase-space representation together
with the corresponding formula: 
\begin{eqnarray*}
{\rm Tr}\exp \left( -\beta H\right) &=&\int d^2\psi \exp \left( -\beta
F(\psi )\right) ,\  \\
\exp \left( -\beta F(\psi )\right) &\equiv &2\left[ \exp \left( -\beta
H\right) \right] _W\left( \psi \right) ,\ \psi \in {\bf C^1},\ 
\end{eqnarray*}
where $\left[ \cdot \cdot \cdot \right] _W$ denotes the Wigner--Weyl symbol
of the corresponding operator (in ${\cal L}^2$),-- and with identifying $%
F(\psi )$ as the Ginzbirg--Landau free energy, we obtain $F(\psi
)=F_0+A\left| \psi \right| ^2+\frac 12B\left| \psi \right| ^4+\cdot \cdot
\cdot ,$ where $F_0,A,B,...$ are explicitly given if there is given the
operator $H$ . Inversely, by a given free energy $F(\psi )$ we obtain $%
H=H\left( N\right) =H_0-\beta ^{-1}\sum_{(n)}\left( \left( -1\right)
^n/n!\right) K_nN^n$, where the coefficients are explicitly given if there
is given the function $F(\psi )$; $\{K_n\}$ have the structure of cumulants.
Note, that evaluation procedure of this point is of interest in its own
right.

Let us now return back to model of the biological junction together with the
number of $Q$ carriers as the distinctive variable, and start from the
problem of two sides, $1$ and $2$, of a synapse coupled by this junction.
Firstly, if we neglect the coupling between sides $1$ and $2$, the
Hamiltonian operator breaks into two parts $H_1+H_2$ . Further, if no
external voltage is applied on the junction, the chemical potential on the
sides $1$ and $2$ are equal. It means that between the states $\left(
N_1,N_2\right) $ and $\left( N_1-\nu ,N_2+\nu \right) $, $\nu \in {\bf Z}$ ,
no difference.

Let us now allow a coupling between $1$ and $2$. It can be splited into two
parts: (1) electrostatic, with a capacity $C$; (2) a charge passing from one
side to other with the extracting, for example, of $Q$ in $1$ and bringing
it in $2$ .

Let us now realize the state spaces of $1$ and $2$ as the Hardy spaces {\bf H%
}$^2$ and define in these spaces the pairs of the number-phase operators $%
\left( N_1,\Phi _1\right) $ and $\left( N_2,\Phi _2\right) $ , $\left[ \Phi
_k,N_l\right] =i\delta _{kl}$, $k,l=1,2;$ together with the Toeplitz
partially isometric one-sided translation operators $T_{{%
{+  \atop -}%
}}^1$, $T_{{%
{+  \atop -}%
}}^2$ . And let $\left\{ e_n^{\left( 1\right) }\right\} $ and $\left\{
e_n^{\left( 2\right) }\right\} $ are the standard basis in {\bf H}$_1^2$ and 
{\bf H}$_2^2$ correspondingly. In this case we can construct the state space
as ${\cal H=}${\bf H}$_1^2\otimes ${\bf H}$_2^2$, and define the following
operators 
\begin{eqnarray*}
N_0 &=&N_1\otimes 1+1\otimes N_2,\quad N=N_1\otimes 1-1\otimes N_2, \\
\quad \Phi &=&\frac 12\left( \Phi _1\otimes 1-1\otimes \Phi _2\right) ,
\end{eqnarray*}
with the commutation relations 
\[
\left[ \Phi ,N_0\right] =0,\quad \left[ \Phi ,N\right] =i, 
\]
on a dense domain in ${\cal H}$. We suppose also that $N_0$ is fixed. The
pair $\left( N,\Phi \right) $ is principal set of operators. It is easy to
see that it is convenient to take {\bf H}$_{-}^2$ instead of {\bf H}$_2^2$ ,
where {\bf H}$_{-}^2$ is the subspace of {\bf L}$^2={\cal L}^2\left(
C_{1,}d\varphi /2\pi \right) $ spanned on $\left\{ e_{-n}\right\} _0^\infty $%
. It implies some evident overdetermination.

At a given $N_0$, $N$ can takes $\left( 2N_0+1\right) $ values. Under this
condition we can explicitly realize ${\cal H}$ as the subspace of the
Laurent space ${\bf L}^2$. In this way we will be prepared to take down a
junction Hamiltonian explicitly: 
\[
H=H_0+\frac 1{2C}N^2+T, 
\]
where $C$ is electric capacity and $T$ is operator of extracting a charge in 
$1$ and bringing it in $2$: $T=t_1\left( T_{+}+T_{-}\right) +t_2\left(
T_{+}^2+T_{-}^2\right) +...$ , -- most probably 
\[
T=t\left( T_{+}+T_{-}\right) , 
\]
and we can choose $t$ real.

As first test of the model presented we can make passage to ``classical''
limit of the corresponding equations of motion. Note, that using the Wigner
phase-space representation as well as the coherent states representation we
obtain the equations are analogous to known Josephson equations.

\medskip

In conclusion, we hope that the model has enough wealth of detail relevant
to biological insight as well as interesting physics. But that is all for
this primary presentation.

\medskip \medskip

\noindent{\bf Acknowledgments}\quad This work was supported in part by
National Foundation for Basic Research (Grant No. F4/310-97).

\medskip

\begin{center}
------------------------------------------------------------
\end{center}

\begin{description}
\item[{\lbrack 1]}]  {\small J. Lubke and oths, J. of Neuroscience 16, 3209
(1996).}

\item[{\lbrack 2]}]  {\small G.M. Sheppard, Neurobiology (Oxford Univ.
Press, New York, 1983). }

\item[{\lbrack 3]}]  {\small K. Hoffman, Banach Spaces of Analytical
Functions (Prentice-Hall, Englewood Cliffs, 1962).}
\end{description}

\end{document}